\documentclass[
aps,
prl,
floatfix,
twocolumn
]{revtex4-1}
\usepackage{graphicx}
\usepackage{amsmath,amssymb}
\usepackage{float}
\usepackage{color}
\usepackage{graphics}
\usepackage{epsfig}

\makeatletter
\@ifundefined{textcolor}{}
{
	\definecolor{BLACK}{gray}{0}
	\definecolor{WHITE}{gray}{1}
	\definecolor{RED}{rgb}{1,0,0}
	\definecolor{GREEN}{rgb}{0,1,0}
	\definecolor{BLUE}{rgb}{0,0,1}
	\definecolor{CYAN}{cmyk}{1,0,0,0}
	\definecolor{MAGENTA}{cmyk}{0,1,0,0}
	\definecolor{YELLOW}{cmyk}{0,0,1,0}
}

\makeatother

\begin{document}

\title{Transverse instability of dark solitons in spin-orbit coupled polariton condensates}
\author{Dmitry A. Zezyulin$^1$}
\author{Yaroslav V. Kartasov$^{2,3}$}

\affiliation{$^1$ITMO University, St.~Petersburg 197101, Russia}

\affiliation{$^2$ICFO-Institut de Ciencies Fotoniques, The Barcelona Institute of Science and Technology, 08860 Castelldefels (Barcelona), Spain}

\affiliation{$^3$Institute of Spectroscopy, Russian Academy of Sciences, Troitsk, Moscow, 108840, Russia}

\begin{abstract}
We consider  dark solitons and their stability in spin-orbit coupled polariton condensates. The system supports spinor solitons of two types: conventional (symmetric) dark solitons and asymmetric half-dark solitons. They demonstrate essentially different behavior upon variation of the strength of spin-orbit coupling. One-dimensional spin-orbit coupled dark solitons are usually unstable, while half-dark solitons can be stable. Two-dimensional dark solitons at early stages of the development of transverse instabilities turn into asymmetric snaking patterns and later into sets of vortex-antivortex solitons with notably different shapes. Depending on the sign of spin-orbit coupling two distinct instability scenarios are possible for such solitons, in which vortices in one component correspond to vortices or antivortices in other component. The decay of two-dimensional half-dark solitons results in the formation of half-vortex chains.
\end{abstract}

\maketitle

\textit{Introduction.} Dark solitons are the fundamental soliton solutions of the nonlinear Schr\"odinger equation, which exist when the product of coefficients determining dispersion and nonlinearity strength is negative \cite{dark01}. One-dimensional (1D) dark soliton represents a density dip residing on the modulationally stable background. Dark solitons were observed in diverse systems, including optical fibers \cite{dark02,dark03} and Bose-Einstein condensates \cite{dark04}. Being stable entities in the 1D  case, dark solitons in 2D usually suffer from transverse ``snake'' instabilities \cite{dark05}, leading to their decay into sets of vortex-antivortex solitons observed experimentally \cite{dark06,dark07}, see also reviews \cite{dark08,dark09}. In the case, when dark solitons are generated in supersonic flows past localized obstacles, their snake instability turns into convective one \cite{dark10}. Such solitons elongate faster than they are destroyed by the instability, that allowed their observation in polariton condensates \cite{dark11}.

Polariton condensates forming in planar microcavities and exhibiting strong nonlinear effects due to repulsive exciton-exciton interactions, represent an ideal platform for the exploration of physics of dark and vortex solitons \cite{dark12}. 
Polariton condensates 
support formation of oblique dark \cite{dark11} and bright \cite{dark13} solitons, nonequilibrium dark solitons supported by resonant \cite{dark14,resn01} and nonresonant \cite{nonr01,nonr02,nonr03,nonr04} pump, spontaneous formation \cite{dark15} and nucleation \cite{dark16,dark17,dark18} of vortices in the flow past obstacles, etc. Polarization phenomena arising from spin degree of freedom substantially enrich evolution of condensates \cite{dark19} leading to new types of spinor solitons, such as half-dark \cite{dark20,dark21} and half-vortex solitons \cite{dark22}. Inclusion of polarization effects requires taking into account TE-TM splitting of polariton energy levels, that can be interpreted as an effective spin-orbit coupling (SOC) \cite{dark23,dark24}. This coupling
dramatically changes properties and the very structure of 
polariton solitons \cite{dark25}.

The effect of SOC was discussed upon generation of oblique half-dark solitons \cite{dark20}, but its influence on stability of dark polariton solitons remains practically unexplored. Thus, in \cite{nonr02} a reduced to usual linear coupling form of SOC was considered. The only work \cite{nonr04} dealing with decay of two-dimensional dark polariton solitons did not take into account polarization effects. The impact of SOC on stability of dark solitons was considered in Bose-Einstein condensates \cite{dark26,dark27} and optical waveguides \cite{dark28}, where this coupling has completely different form and physical origin than coupling in polaritonic systems.

The aim of this Letter is to show that even small SOC drastically affects the development of the transverse instability (TI) of dark polariton solitons, leading to  completely different dynamical patterns from those 
for scalar dark solitons. They are composed from vortex-antivortex pairs, whose disposition depends on the sign of SOC. Due to SOC, vortices and antivortices in emerging pairs acquire different shapes. Two different regimes are encountered, when for positive/negative coupling strength the appearance of vortex in one spin component is followed by the formation of antivortex/vortex in other component. We also found that half-dark solitons split into half-dark vortex pairs.

\textit{The model.} We consider spin-orbit coupled polariton condensate in a 2D  microcavity. We are interested in quasi-conservative dynamics governed by the normalized  equations \citep{dark25}:
\begin{equation*}
\label{eq:main}
\begin{array}{l}
i \partial_t{\psi}_\pm = \left[-\frac{1}{2}\nabla^2 + |\psi_\pm|^2 + \sigma|\psi_\mp|^2\right]\psi_\pm
+\beta(\partial_x \mp i\partial_y)^2\psi_\mp,
\end{array}
\end{equation*}
In the experiment, the unavoidable losses are expected to affect the  polariton dynamics only quantitatively (see e.g. \cite{lifetime01}), e.g. one may expect gradual shrinkage of instability band due to decrease of peak amplitude. According to the recent experiments \cite{lifetime02}, the polariton lifetime can be increased to hundreds of picoseconds in high-quality microcavity samples that substantially exceeds time intervals considered below. $\boldsymbol{\psi}(x,y,t)=(\psi_+, \psi_-)^\textrm{T}$ is the spinor wavefunction in circular polarization basis; $\partial_{t,x,y}$ are the partial derivatives; $\nabla^2 = \partial_{x}^2 + \partial_{y}^2$ is the Laplacian; time $t$ and spatial coordinates $x,y$ are measured in units which ensure $\hbar=1$ and $m^*=1$, where $m^*$ is the effective polariton mass. The model accounts for repulsive interactions of polaritons with the same spin, while small $\sigma<0$ characterizes the attraction of polaritons with opposite spins (we set $\sigma=-0.05$, that is the experimentally relevant value); $\beta$ is the SOC strength proportional to the difference of effective masses of TE and TM polaritons.

\begin{figure}
\begin{center}		\includegraphics[width=\columnwidth]{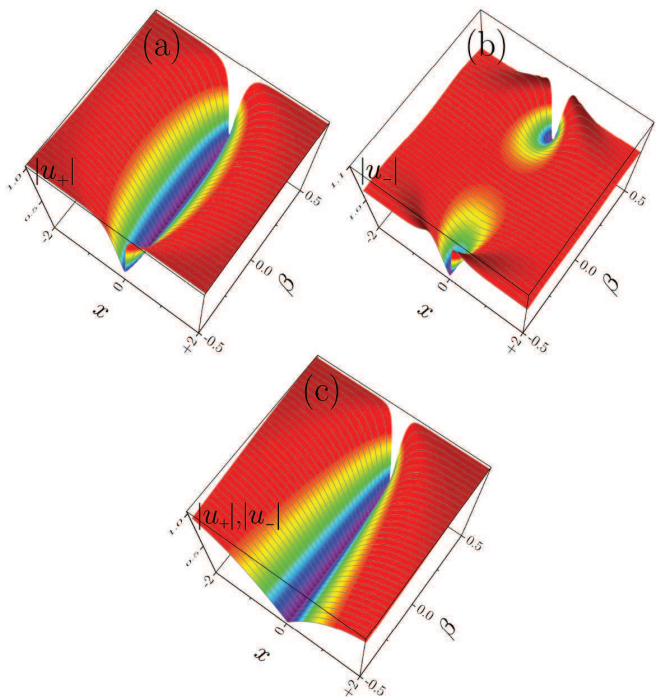}
\caption{Modification of the shape of $u_+$ (a) and $u_-$ (b) components of half‐dark soliton and of $u_\pm$ components (c) of dark soliton upon variation of SOC strength $\beta$. Here and in figures below $\mu=1$ and $\sigma=-0.05$. \label{fig:profiles}}
\end{center}
\end{figure} 
 
 \textit{Soliton solutions.} We look for  stationary quasi-one-dimensional solutions that depend only on one spatial coordinate $x$ and are uniform in $y$. Using the substitution  $\psi_\pm = e^{-i\mu t} u_\pm(x)$,
where $\mu$ is the chemical potential, and solving the resulting system for $u_\pm(x)$, we have encountered  solutions of two types. Solution of the first type is obtained using simple reduction in the form $u_-(x) = \pm u_+(x)$, i.e., it represents either symmetric (upper sign) or antisymmetric (lower sign) state, where $u_+$ is the conventional $\tanh$-shaped dark soliton:

\begin{equation}
\label{eq:ds}
u_+(x) = \sqrt{\frac{\mu}{1+\sigma}}\tanh\left(\sqrt{\frac{\mu}{1\mp2\beta}}\ x\right),
\end{equation}
where sign in the denominator should be chosen according to type of solution. The properties (including stability) of symmetric and antisymmetric modes are similar (the properties of one family are obtained from those of other family by the inversion of $\beta$ sign), so we focus on symmetric modes. The solution (\ref{eq:ds}) immediately shows that the increase of $\beta$ from negative to positive values leads to soliton steepening, while in the limit $\beta\to 1/2$ (which however may not be practically reachable) the soliton becomes infinitely narrow for a fixed $\mu$, see Fig.~\ref{fig:profiles}(c).

Besides conventional dark solitons, the system also supports asymmetric \textit{half-dark} solitons with significantly different shapes of the two components. The latter can be found numerically and their transformation with $\beta$ is illustrated in Figs.~\ref{fig:profiles}(a,b). 
For small $\beta$ values half-dark solitons feature a prominent density dip only in the $u_+$ component, with $u_-$ component being nearly uniform. The increase of SOC strength to larger positive or negative values leads to steepening of the $u_+$ component and development of a dip surrounded by two maxima in the $u_-$ component. Remarkably, for half-dark solitons inversion of the sign of $\beta$ does not change the shape of solution. This is in clear contrast to usual dark solitons that have different widths for positive and negative $\beta$ values, see Fig.~\ref{fig:profiles}(c). This suggests that stability properties of half-dark solitons depend on $|\beta|$ only, while for dark solitons the sign of $\beta$ is important too.

\begin{figure}
\begin{center}		\includegraphics[width=\columnwidth]{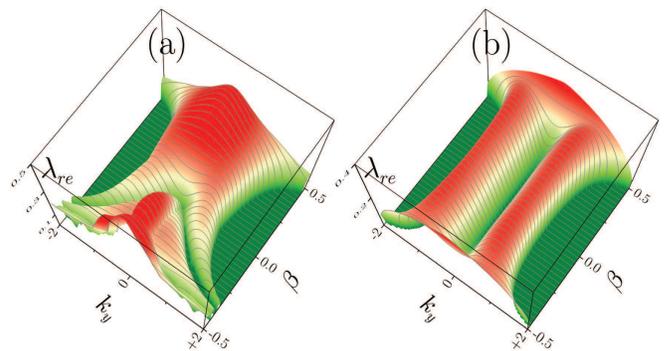}
\caption{Maximal instability increment as a function of modulation frequency $k_y$ and SOC strength $\beta$ for dark (a) and half‐dark (b) solitons. 
\label{fig:growth}}
\end{center}
\end{figure} 

\begin{figure}
\includegraphics[width=0.99\columnwidth]{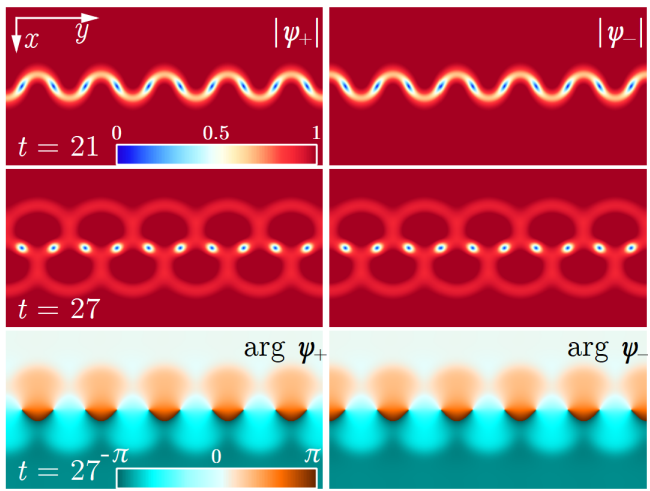}
\caption{Decay of the dark soliton at $\beta=0$. $|\psi_+|$ and $|\psi_-|$ components are shown in the left and right columns, respectively. For $t=27$ we also show phases $\textrm{arg}\,\psi_\pm$. All distributions are shown within $(x,y) \in[-15,15]\times  [-33.4,33.4]$ windows. See detailed dynamics in \textcolor{blue}{Visualization 1}. \label{fig:ev30}}
\end{figure}
    
\textit{Transverse instability.} 
First, we address the spectral instability of dark solitons with respect to small-amplitude two-dimensional perturbations. We consider perturbed solutions in the form $\psi_\pm = e^{-i\mu t}[u_\pm(x) + 
p_\pm(x)e^{ik_yy+\lambda t} + q_\pm^*(x)e^{-ik_yy+\lambda^* t}]$, where $p_\pm(x)$ and $q_\pm(x)$ are small perturbations localized in $x$, $k_y$ is the spatial frequency of perturbation in the $y$ direction, and $\lambda=\lambda_{re} + i\lambda_{im}$ is a complex eigenvalue, whose real part $\lambda_{re}$ determines the instability  growth rate. Linearization 
around stationary solution $u_{\pm}$ leads to the eigenvalue problem for $\lambda$, which can be solved numerically for different values of SOC strength $\beta$ and frequency $k_y$.  

\begin{figure}[!t]
\includegraphics[width=0.98\columnwidth]{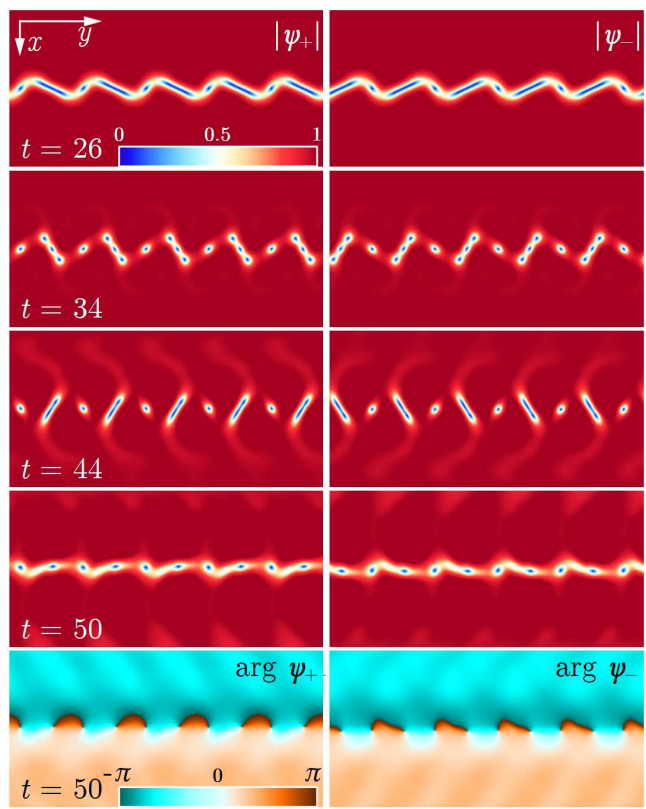}
\caption{Decay of the dark soliton at $\beta=-0.1$. $\psi_+$ and $\psi_-$ are shown in left and right columns, respectively, within $(x,y) \in [-15,15]
   \times  [-33.4,33.4]$ windows. For $t=50$ both amplitudes and phases are shown. See \textcolor{blue}{Visualization 2}. \label{fig:ev22}}
\end{figure}
\begin{figure}[!t]
\includegraphics[width=0.98\columnwidth]{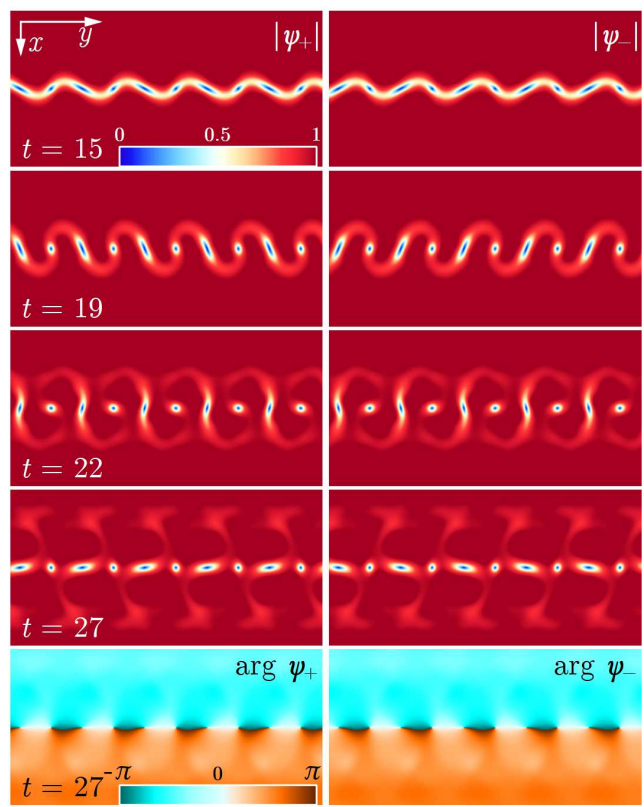}
\caption{Decay of dark soliton at $\beta=0.1$. $\psi_+$ and $\psi_-$ are shown in left and right columns, respectively, within $(x,y) \in [-15,15]\times [-33.4,33.4]$ windows. For $t=27$ both amplitudes and phases are shown. See \textcolor{blue}{Visualization 3}. \label{fig:ev23}}
\end{figure}

It is well known \cite{dark05} for  $\beta=\sigma=0$ that usual dark solitons are  unstable for the modulation frequencies within the interval $0<|k_y|^2<1$. This situation  changes significantly when SOC strength becomes nonzero, and weak cross-attraction $\sigma$ is taken into account. In Fig.~\ref{fig:growth}(a) we  plot  the maximal instability increment computed for dark solitons with representative chemical potential $\mu=1$
on the plane of parameters $(k_y,\beta)$. Notice that instability band is asymmetric in $\beta$ as it was supposed above. It notably expands with increase of $|\beta|$. Fig.~\ref{fig:growth}(a) shows that even weak attraction of cross-polarized polaritons leads to the 1D instability of dark solitons at $k_y=0$. Moreover, the increase of SOC strength $\beta$ to large (positive or negative) values clearly enhances one-dimensional instability. For $|\beta|>0.1$ the instability associated with $k_y=0$ is the strongest one, i.e. the corresponding increment is maximal among all $k_y$. Thus   for large   $|\beta|$, if instability is seeded by noisy perturbations, the snaking may not occur and dynamics will be ruled mostly by the 1D instability. Therefore, proceeding to investigation of dynamical development of TI, we limit the consideration to relatively small values of $\beta$. In Figs.~\ref{fig:ev30}, \ref{fig:ev22} and \ref{fig:ev23} we illustrate the development of TI seeded by 
small initial perturbations with the same frequency $k_y\approx  0.5$ for three representative values of $\beta$. In the absence of SOC ($\beta=0$, Fig.~\ref{fig:ev30}), TI results  in the development of symmetric snaking pattern (identical in both components) from input dark soliton stripe, which then breaks into alternating vortices and antivortices, i.e.,  topological defects with winding numbers equal to $+1$ and $-1$. For nonzero $\beta$ (Fig.~ \ref{fig:ev22} and \ref{fig:ev23}) the emerging snaking patterns are strongly asymmetric and different in two components, they break into segments of unequal lengths, such that longer segments rotate in the $(x,y)$ plane and direction of rotation is opposite in the $\psi_+$ and $\psi_-$ components. Notice the appearance of elements with three close vortices that alternate with antivortices in Fig.~\ref{fig:ev22} that was never observed in scalar systems. Strong asymmetry in shapes of vortices and antivortices in the chain due to SOC is a distinctive feature of this system. At later stages of the instability development, one observes the formation of vortex-antivortex pairs. At the same time, the comparison of lower panels of Figs.~\ref{fig:ev22} and \ref{fig:ev23} reveals qualitative difference of phase patterns for opposite signs of SOC. Indeed, in Fig.~\ref{fig:ev22} the position of each vortex in $\psi_+$ component approximately coincides with a vortex in $\psi_-$ component, and each antivortex in $\psi_+$ corresponds to an antivortex in $\psi_-$. In Fig.~\ref{fig:ev23} one observes completely different picture where each vortex in $\psi_+$ component corresponds to an antivortex in $\psi_-$, and, \textit{vice versa}, each antivortex in $\psi_+$ corresponds to a vortex in $\psi_-$. This phenomenon does not exist in scalar systems.

\begin{figure}
\includegraphics[width=0.99\columnwidth]{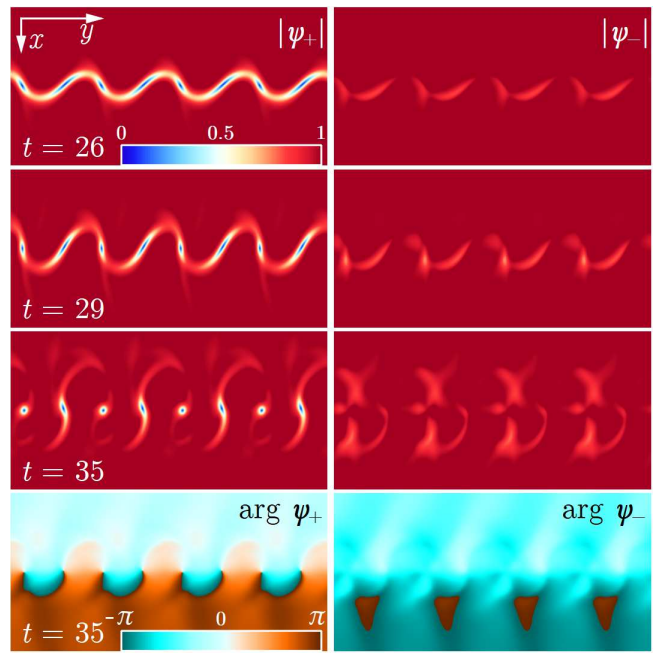}
	\caption{Decay of the half-dark soliton with $\beta=0.4$. $\psi_+$ and $\psi_-$ correspond to left and right columns, respectively.   For $t=35$ both amplitudes and phases are shown. All panels are shown within $(x,y) \in [-10,10]\times [-33.4,33.4]$ windows. See \textcolor{blue}{Visualization 4}. \label{fig:ev28}}
\end{figure}

Now we turn to TI of asymmetric half-dark solitons. Their instability increments are plotted in Fig.~\ref{fig:growth}(b) on the $(k_y,\beta)$ plane. The dependence is symmetric in $\beta$. In contrast to dark solitons [Fig.~\ref{fig:growth}(a)], half-dark ones are most unstable with respect to perturbations with nonzero modulation frequency $k_y$ (except for large values of $\beta$ close to $\pm1/2$). Moreover, for small values of SOC strength $|\beta|   <0.2$, 1D  half-dark solitons are completely stable, i.e., $\lambda_{re}=0$ at $k_y=0$. Half-dark solitons demonstrate very unusual TI, dynamics depicted in Fig.~\ref{fig:ev28}, where one again observes the asymmetric snaking pattern, which however is well-pronounced only in the $\psi_+$ component, while $\psi_-$ component features only relatively weak density modulations, even though the SOC strength in this case was relatively large ($\beta=0.4$). Respectively, at later stages of TI development one observes the excitation of the chains of half-vortices (i.e. vortices appear only in $\psi_+$ component), where nearly circularly symmetric states alternate with extremely asymmetric ones.

\textit{To conclude},  SOC significantly affects stability of 1D and 2D dimensional dark and half-dark solitons in polariton condensates. SOC leads to unusual asymmetric snaking pattern at early stages of TI development, and at later stages  solutions break into vortex-antivortex pairs such that the resulting phase distribution can be distinctively different for opposite signs of SOC. 

\section*{Funding Information}
Government of Russian Federation (Grant 08-08).  Ministry of Education and Science of Russian Federation (Megagrant No. 14.Y26.31.0015). Severo Ochoa program (SEV-2015-0522) of the Government of Spain, Fundacio Cellex, Fundació Mir-Puig.



\begin{thebibliography}{1}

\bibitem{dark01} V. E. Zakharov and A. B. Shabat, "Exact theory of two-dimensional self-focusing and one-dimensional self-modulation of wave in nonlinear media," Sov. Phys. J. Exp. Theor. Phys. \textbf{34}, 62 (1972).

\bibitem{dark02} P. Emplit, J. P. Hamaide, R. Reynard, C. Froehly, and A. Barthelemy, "Picosecond steps and dark pulses through nonlinear single-mode fibers," Opt. Commun. \textbf{62}, 374 (1987).

\bibitem{dark03} D. Krokel, N. J. Halas, G. Giuliani, and D. Grischkowsky, “Dark-pulse propagation in optical fibers", Phys. Rev. Lett. \textbf{60}, 29 (1988).

\bibitem{dark04} S. Burger, K. Bongs, S. Dettmer,W. Ertmer, K. Sengstock, A. Sanpera, G. V. Shlyapnikov, and M. Lewenstein, "Dark solitons in Bose-Einstein condensates," Phys. Rev. Lett. \textbf{83}, 5198 (1999).

\bibitem{dark05}  E. A.  Kuznetsov,  S. K. Turitsyn, "Instability and collapse of solitons in media with a defocusing nonlinearity," Sov. Phys. J. Exp. Theor. Phys. \textbf{67},  1583 (1988).

\bibitem{dark06} A. V. Mamaev, M. Saffman, and A. A. Zozulya, "Propagation of dark stripe beams in nonlinear media: Snake instability and creation of optical vortices," Phys. Rev. Lett. \textbf{76}, 2262 (1996).

\bibitem{dark07} Z. Dutton, M. Budde, C. Slowe, and L. V. Hau, "Observation of quantum shock waves created with ultra-compressed slow light pulses in a Bose-Einstein condensate," Science \textbf{293}, 663 (2001).

\bibitem{dark08} Y. S. Kivshar and B. Luther-Davies, "Dark optical solitons: physics and applications," Phys. Rep. \textbf{298}, 81 (1998).

\bibitem{dark09} Y. S. Kivshar and D. E. Pelinovsky, "Self-focusing and transverse instabilities of solitary waves," Phys. Rep. \textbf{331}, 117 (2000).

\bibitem{dark10} A. M. Kamchatnov and L. P. Pitaevskii, "Stabilization of solitons generated by a supersonic flow of Bose-Einstein condensate past an obstacle," Phys. Rev. Lett. \textbf{100}, 160402 (2008).

\bibitem{dark11} A. Amo, S. Pigeon, D. Sunvitto, V. G. Sala, R. Hivet, I. Carusotto, F. Pisanello, G. Lemenager, R. Houdre, E. Giacobino, C. Ciuti, and A. Bramati, "Polariton superfluids reveal quantum hydrodynamic solitons," Science \textbf{332}, 1167 (2011).

\bibitem{dark12} I. Carusotto and C. Ciuti, "Quantum fluids of light," Rev. Mod. Phys. \textbf{85}, 299 (2013).

\bibitem{dark13} M. Sich, D. N. Krizhanovskii, M. S. Skolnick, A. V. Gorbach, R. Hartley, D. V. Skryabin, E. A. Cerda-Mendez, K. Biermann, R. Hey, and P. V. Santos, "Observation of bright solitons in a semiconductor microcavity," Nat. Photon. \textbf{6}, 50 (2012).

\bibitem{dark14} A. V. Yulin, O. A. Egorov, F. Lederer, and D. V. Skryabin, "Dark polariton solitons in semiconductor microcavities," Phys. Rev. A \textbf{78}, 061801(R) (2008).

\bibitem{resn01} V. Goblot, H. S. Nguyen, I. Carusotto, E. Galopin, A. Lemaître, I. Sagnes, A. Amo, and J. Bloch, "Phase-controlled bistability of a dark soliton train in a polariton fluid," Phys. Rev. Lett. \textbf{117}, 217401 (2016).

\bibitem{nonr01} Y. Xue and M. Matuszewski, "Creation and abrupt decay of a quasistationary dark soliton in a polariton condensate," Phys. Rev. Lett. \textbf{112}, 216401 (2014).

\bibitem{nonr02} F. Pinsker and H. Flayac, "On-demand dark soliton train manipulation in a spinor polariton condensate," Phys. Rev. Lett. \textbf{112}, 140405 (2014).

\bibitem{nonr03} T.-W. Chen, S.-D. Jheng, W.-F. Hsieh, S.-C. Cheng, "Dark solitons in a spinor microcavity-polariton condensate," Superlattices Microstruct.  \textbf{98},  96 (2016).

\bibitem{nonr04} L. A. Smirnov, D. A. Smirnova,  E. A. Ostrovskaya, and Yu. S. Kivshar, "Dynamics and stability of dark solitons in exciton-polariton condensates," Phys. Rev. B  \textbf{89}, 235310 (2014).

\bibitem{dark15} K. G.  Lagoudakis, M. Wouters, M. Richard, A. Baas, I. Carusotto, R. André, Le Si Dang, and B. Deveaud Plédran, "Quantized vortices in an exciton–polariton condensate," Nat. Phys. \textbf{4}, 706 (2008)

\bibitem{dark16} G. Nardin, G. Grosso, Y. Leger, B. Pietka, F. Morier-Genoud, D. Deveaud-Pledran, "Hydrodynamic nucleation of quantized vortex pairs in a polariton quantum fluid," Nat. Phys. \textbf{7}, 635 (2011).

\bibitem{dark17} S. Pigeon, I. Carusotto, and C. Ciuti, "Hydrodynamic nucleation of vortices and solitons in a resonantly excited polariton superfluid," Phys. Rev. B \textbf{83}, 144513 (2011).

\bibitem{dark18} G. Grosso, G. Nardin, F. Morier-Genoud, Y. Leger, and B. Deveaud-Pledran, "Soliton instabilities and vortex street formation in a polariton quantum fluid," Phys. Rev. Lett. \textbf{107}, 245301 (2011).

\bibitem{dark19} A. Werner, O. A. Egorov, and F. Lederer, "Spin dynamics of dark polariton solitons," Phys. Rev. B \textbf{85}, 115315 (2012).

\bibitem{dark20} H. Flayac, D. D. Solnyshkov, and G. Malpuech, "Oblique half-solitons and their generation in exciton-polariton condensates," Phys. Rev. B \textbf{83}, 193305 (2011).

\bibitem{dark21} R. Hivet, H. Flayac, D. D. Solnyshkov, D. Tanese, T. Boulier, D. Andreoli, E. Giacobino, J. Bloch, A. Bramati, G. Malpuech, and A. Amo, "Half-solitons in a polariton quantum fluid behave like magnetic monopoles," Nat. Phys. 8, 724 (2012).

\bibitem{dark22} K. G. Lagoudakis, T. Ostatnický, A. V. Kavokin, Y. G. Rubo, R. André, B. Deveaud-Plédran, "Observation of half-quantum vortices in an exciton-polariton condensate," Science \textbf{326}, 974 (2009).

\bibitem{dark23} V. G. Sala, D. D. Solnyshkov, I. Carusotto, T. Jacqmin, A. Lemaitre, H. Terças, A. Nalitov, M. Abbarchi, E. Galopin, I. Sagnes, J. Bloch, G. Malpuech, and A. Amo, "Spin-orbit coupling for photons and polaritons in microstructures," Phys. Rev. X \textbf{5}, 011034 (2015).

\bibitem{dark24} S. Dufferwiel, F. Li, E. Cancellieri, L. Giriunas, A. A. P. Trichet, D. M. Whittaker, P. M. Walker, F. Fras, E. Clarke, J. M. Smith, M. S. Skolnick, D. N. Krizhanovskii, "Spin textures of exciton-polaritons in a tunable microcavity with large TE-TM splitting," Phys. Rev. Lett. \textbf{115}, 246401 (2015).

\bibitem{dark25} H. Flayac, I. A. Shelykh, D. D. Solnyshkov, and G. Malpuech, "Topological stability of the half-vortices in spinor exciton-polariton condensates," Phys. Rev. B \textbf{81}, 045318 (2010).

\bibitem{dark26} V. Achilleos, J. Stockhofe, P. G. Kevrekidis, D. J. Frantzeskakis, and P. Schmelcher, "Matter-wave dark solitons and their excitation spectra in spin-orbit coupled Bose-Einstein condensates," Europhys. Lett. \textbf{103}, 20002 (2013).

\bibitem{dark27} Y. V. Kartashov and A. M. Kamchatnov, "Wave patterns generated by a flow of a two-component Bose-Einstein condensate with spin-orbit interaction past a localized obstacle,"  Europhys. Lett. \textbf{107}, 10008 (2014).

\bibitem{dark28} Y. V. Kartashov, V. V. Konotop, and B. A. Malomed, "Dark solitons in dual-core waveguides with
dispersive coupling," Opt. Lett. \textbf{40}, 4126 (2015).

\bibitem{lifetime01}
 M. Walker, L. Tinkler, D. V. Skryabin, A. Yulin, B. Royall, I. Farrer, D. A. Ritchie, M. S. Skolnik, D. N.  Krizhanovskii,  
 ``Ultra-low-power hybrid light-matter solitons.''
 Nat.
Commun. {\bf 6}, 8317 (2015).

\bibitem{lifetime02}  Y. Sun, P. Wen, Y.  Yoon, G. Liu, M. Steger, L. N. Pfeiffer, K. West, D. W.  Snoke, K. A. Nelson,
``Bose-Einstein Condensation of Long-Lifetime Polaritons in Thermal Equilibrium.'' 
Phys. Rev. Lett. {\bf 118}, 016602 (2017).

\end{thebibliography}
\end{document}